\documentclass[twocolumn,prl,showpacs]{revtex4}
\usepackage{graphicx,color}
\usepackage{dcolumn}
\begin{document}
%
\title{
 Phase Diagram of the Square-Lattice Three-State Potts Antiferromagnet
 with Staggered Polarization Field
 }
\author{
 Hiromi Otsuka
 and 
 Yutaka Okabe
}
\address{
 Department of Physics, Tokyo Metropolitan University,
 Tokyo 192-0397 Japan
}
\date{\today}
\begin{abstract}
 We study a square-lattice three-state Potts antiferromagnet with a
 staggered polarization field at finite temperature.
 Numerically treating the transfer matrices, we determine two phase
 boundaries separating the model-parameter space into three parts.
 We confirm that one of them belongs to the ferromagnetic three-state
 Potts criticality, which is in accord with a recent prediction, and
 another to the Ising type; these are both corresponding to the massless 
 renormalization-group flows stemming from the Gaussian fixed points.
 We also discuss a field theory to describe the latter Ising transition.
\end{abstract}
\maketitle


 It is widely recognized that strong frustrations can provide a ground
 state with an extensive entropy and prevent long-range orderings of
 systems.
 The simplest example may be the triangular-lattice antiferromagnetic
 (AF) Ising model, whose ground state is critical and exhibits power-law
 decays of correlation functions
 \cite{Wann50,Hout50}.
 Although, unlike in the ferromagnetic (F) cases, the models of this
 type may depend on their details (e.g., lattice structures), the field
 theories used to describe the ground state and lower-energy excitations
 have attracted much attention.
 Further, recent interests in this area are rather focused on mutual
 relations of fixed points (e.g., crossovers of criticalities) embedded
 in the renormalization-group (RG) flows
 \cite{Alex75},
 so the understandings of their universal properties are quite important
 \cite{Zamo86}.

 The AF three-state ($q=3$) Potts model on the square lattice $\Lambda$ 
 exhibits the same properties as those systems; its Hamiltonian is
 described by using the ternary variables $\sigma_j=0,1,2$
 ($j\in\Lambda$) as
 \begin{equation}
  H_0=J\sum_{\langle j,k\rangle} \delta_{\sigma_j,\sigma_{k}}~~~~(J>0),
   \label{eq_Hami0}
 \end{equation}
 where the sum runs over all nearest-neighboring (NN) pairs.
 While its ground-state
 \cite{Lena67,Baxt70,Nijs82,Burt97,Delf01,Kola84,Sala98}
 and finite temperature properties
 \cite{Card01,Ferr99}
 have been intensively investigated, a possibility of the
 crossover behavior from the AF to the F three-state Potts criticality
 has been recently proposed on the basis of a continuous field theory
 \cite{DelfNATO}.
 In close relation to the classical spin system, the realization of
 the field theory was found in the 1D quantum spin system (i.e., a
 frustrated XXZ Heisenberg chain in magnetic fields), and it can show
 the F three-state Potts criticality in its ground state
 \cite{Lech03}. 
 In this Letter, we quantitatively investigate the possibility of
 crossover by introducing a square-lattice three-state Potts model
 defined below, which is also relevant to 1D quantum systems.
 Before presenting the formulation of our investigation, we shall
 briefly refer to the relating research so far.

 Since the ground state of $H_0$ is equivalent to the six-vertex model
 on the ice point, it shows the Gaussian criticality with the conformal
 anomaly number $c=1$ 
 \cite{Lena67,Baxt70,Nijs82,Burt97,Delf01}.
 This sort of equivalence was used to determine the properties
 of lower-energy excitations: 
 den Nijs, Nightingale, and Schick
 found that both the uniform
 $s_j=e^{2\pi i \sigma_j/3}$
 and the staggered magnetizations
 $S_j=(-1)^js_j$
 are relevant with scaling dimensions
 $x_s=\frac23$
 and
 $x_S=\frac16$,
 respectively
 \cite{Nijs82}.
 Here, $(-1)^j=\pm1$ for $j$ in the even (odd) sublattice $\Lambda_\pm$.
 It was pointed out that the scaling dimensions of relevant scalar
 operators can take three values, and other than above two, the
 staggered polarization
 $P_j=(-1)^j\sum'_{k}(2\delta_{\sigma_j,\sigma_{k}}-1)$
 takes
 $x_P=\frac32$
 [the sum is over $k$ next-nearest neighboring (NNN) with $j$]
 \cite{Burt97}. 

 Numerical studies have been also performed
 \cite{Kola84,Sala98}:
 Salas and Sokal using the Monte Carlo (MC) simulations at zero
 temperature succeeded in confirming the lowest two scaling dimensions,
 but the estimation for the third one exhibits a deviation due to its
 larger energy scale
 \cite{Sala98}.
 The scaling dimensions can be also obtained by the transfer matrix
 technique; the lowest two were accurately obtained by de Queiroz, but
 the third could not be found
 \cite{Quei02}.
 More profound understanding at finite temperature was brought about by
 Cardy, Jackobsen, and Sokal
 \cite{Card01}.
 They pointed out that there are two types of excitations controlled by
 the thermal scaling field $u=e^{-J/k_{\rm B}T}$, i.e., the relevant one
 with
 $x_\epsilon=\frac32$, 
 and the marginal one, and that both of these are necessary to explain
 the exotic corrections to scaling observed in 
 the MC data \cite{Ferr99}.
 Nevertheless, as its main role, the energy operator
 $\epsilon_j=\sum_{k}\delta_{\sigma_j,\sigma_{k}}$
 (the sum is over $k$ NN with $j$) brings about the second-order phase
 transition with the divergent correlation length $\xi\propto
 u^{-1/(2-x_\epsilon)}$
 \cite{Burt97,Card01}.

 Recently, Delfino argued the ``bosonization'' descriptions of the
 transition and above-mentioned excitations
 \cite{DelfNATO,Delf01}.
 He also provided their discrete symmetry properties, and then concluded
 that the criticality in the ground state can exhibit the crossover
 behavior to the three-state Potts criticality with $c=\frac45$ as a
 resultant of the competing relevant perturbations $\epsilon_j$ and
 $P_j$, and its description is given by the self-dual sine-Gordon model
 with the dimensionless coupling $\beta^2=6\pi$ (see also Ref.\
 [\onlinecite{Lech01}]).


 Based on these developments,
 especially for the aim of quantitative understandings on the
 competitions between relevant perturbations and the resultant
 crossovers of criticalities,
 we shall introduce the following model,
 namely, the square-lattice AF three-state Potts model with the staggered 
 polarization field defined by the reduced Hamiltonian ${\cal
 H}(K,V)=H/k_{\rm B}T$ with
 \begin{equation}
  {\cal H}(K,V)
   =K\sum_{\langle j,k\rangle}       \delta_{\sigma_j,\sigma_{k}}
   -V\sum_{[j,k]}(-1)^j \delta_{\sigma_j,\sigma_{k}}. 
   \label{eq_Hami1}
 \end{equation}
 The second sum runs over all NNN pairs $[j,k]$, so
 that the sublattice symmetry is explicitly broken for nonzero $V$, but
 the ${\bf S}_3$ symmetry associated with the global permutations of the
 ternary variables is preserved.
 Other than $(K,V)=(\infty,0)$, there are two special points in this
 model:
 When $K=0$, the system decouples into those defined on $\Lambda_{\pm}$,
 and thus $(0,V^{\rm ex}_3)$ and $(0,\infty)$ correspond to the
 exact transition point of the F three-state Potts model on $\Lambda_+$
 and to the ground state of the AF three-state Potts model on $\Lambda_-$,
 respectively [$V^{\rm ex}_q:=\ln(1+\sqrt{q})$] \cite{Pott52}.
 Further, when $K>0$, we expect that the disordered phase at $V\ll1$
 will change into a phase with the F ordering on $\Lambda_+$ and with
 the AF ordering on $\Lambda_-$ satisfying the exclusion condition of
 states at $V\gg1$.
 Therefore, we shall numerically clarify the phase diagram in the
 two-dimensional model-parameter space.

 Let us consider $\Lambda$ with $M$ rows of $L$ sites wrapped on a
 cylinder and take an even number $L$ and the limit $M\to\infty$.
 Then, the site $j\in\Lambda$ is specified by $l\in[1,L]$ and
 $m\in[1,M]$.
 For this system, we can define the transfer matrix ${\bf T}(L)$
 connecting the NNN pair of rows, and denote its
 eigenvalues as $\{\lambda_\alpha(L)\}$ ($\alpha$ specifies a level). 
 The conformal field theory provides direct expressions for $c$ and
 $x_\alpha$ (the scaling dimensions) of critical systems by using the
 eigenvalues
 \cite{Card84,Blot86}: 
 \begin{eqnarray}
  && -\zeta\ln\left[\lambda_I(L)\right]\simeq Lf-\pi c/6L+b/L^3,\\
  \label{eq_c}
  && -\zeta\ln\left[\lambda_\alpha(L)\right]
     +\zeta\ln\left[\lambda_I(L)     \right]\simeq 2\pi x_\alpha/L, 
  \label{eq_x}
 \end{eqnarray}
 where $\lambda_I$ is the largest one corresponding to the ground
 state.
 $\zeta$, $f$, and $b$ are the geometric factor ($\frac12$), a free
 energy per site, and a nonuniversal constant, respectively.
 Here, it should be noted that the discrete symmetries of the lattice
 Hamiltonian, e.g.,
 the translation ${\cal T}$
 (e.g., $\sigma_{l,m} \to \sigma_{l+2,m}$), 
 the space inversion ${\cal P}$
 (e.g., $\sigma_{l,m} \to \sigma_{L-l+2,m}$), 
 and the ${\bf S}_3$ symmetry
 [$\sigma_{l,m} \to g(\sigma_{l,m}$) for $g\in {\bf S}_3$]
 are crucial not only for a reduction of computational efforts,
 but also for the proper specification of the level $\alpha$
 \cite{OtsuAP}.
 This can be clearly demonstrated by evaluating the third scaling
 dimension $x_P$ at $(K,V)=(\infty,0)$. 
 According to Ref.\ [\onlinecite{Delf01}], the staggered polarization
 operator is invariant for all $g\in {\bf S}_3$ and has the wave number
 $\pi/a$ so that the corresponding level
 $-\zeta\ln\left[\lambda_P(L)\right]$
 should be found in the subspace specified by these symmetries.
 In Table\ \ref{TAB_I}, we give the numerical data for the scaling
 dimension $x_P(L)$ estimated from Eq.\ (\ref{eq_x}).
 The extrapolated value does not deviate more than $0.1\%$ from the
 theoretical one, $\frac32$, and the level is not the third one
 in the whole space, so the characterization of the excitation levels is
 essentially important (for other exponents, see Ref.\
 [\onlinecite{Quei02}]).

\begin{table}[t]
 \caption{
 The size dependence of the scaling dimension $x_P$. 
 The fitting $x_P(L)=x_P(\infty)+b_1/L^2+b_2/L^4$ is performed using the
 data of $L=$10, 12, and 14.
 } 
 \begin{tabular}{cccccc}
  \hline\hline

           &~~~{$L=8$}~~~&~~~10~~~ &~~~12~~~&~~~14~~~&  $\infty$ \\ 

  \tableline

  $x_P(L)$ &1.6801468~&~1.6065435~&~1.5710939~&~1.5510358~&~1.5007 \\
          
  \hline\hline
 \end{tabular}
 \label{TAB_I}
\end{table}

 Now, according to the prediction of the crossover, it is plausible that
 the critical RG flow starts out of $(K,V)=(\infty,0)$ and arrives at
 the point $(0,V^{\rm ex}_3)$.
 Thus, at this stage, our numerical task is to determine the line
 $V_3(K)$. 
 For this purpose, we shall employ the phenomenological RG (PRG) method
 \cite{Room80}.
 Let us denote the left-hand side of Eq.\ (\ref{eq_x}) as $\Delta
 E_\alpha(K,V,L)$ (i.e., an excitation gap), then we shall numerically
 solve the following PRG equation for a given value of $K$ with respect
 to $V$: $L\Delta E_\alpha(K,V,L)=L'\Delta E_\alpha(K,V,L')$.
 Since this is satisfied by the gap $\Delta E_\alpha(K,V,L)\propto
 1/L$, the obtained value can be regarded as the size-dependent estimate
 of the transition point,
 say $V_3(K,\bar{L})$ [we take $\bar{L}=(L+L')/2$ and $L'=L-2$ in the
 following].
 Further, there are two critical fixed points connected by the RG flow,
 so a relationship between lower-energy excitations on these fixed
 points, namely the ultraviolet-infrared (UV-IR) operator
 correspondence, is quite important for the choice of the excitation
 $\alpha$.
 Along the flow, the conjecture,
 $S       _j \to \tilde s       _j$
 and 
 $S^{\ast}_j \to \tilde s^{\ast}_j$,
 has been proposed
 \cite{Lech01}. 
 $\tilde s_j$ is the magnetic operator on the IR fixed point with
 the three-state Potts criticality, whose scaling dimension is
 $x_{\tilde s}=\frac{2}{15}$.
 Since the excitation $\tilde s_j$ provides the lowest energy level, 
 we shall focus our attention on $\Delta E_{\tilde s}(K,V,L)$ stemming
 from the level of $S_j$ on the UV Gaussian fixed point.

\begin{figure}[t]
 \includegraphics[width=2.8in]{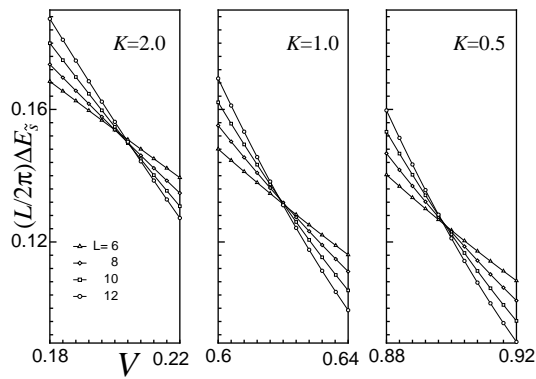}
 \caption{
 The $V$ dependence of $(L/2\pi)\Delta E_{\tilde s}(K,V,L)$.
 From left to right, $K=$ 2.0, 1.0 and 0.5, respectively. 
 The correspondence between marks and system sizes is given in the
 figure. 
 The crossing points give $V_3(K,\bar{L})$.
 }
 \label{FIG1}
\end{figure}

 The exact diagonalization calculations of ${\bf T}(L)$ with $L=$4-14
 are performed by  using the Lanczos algorithm, where the discrete
 symmetries ${\cal T}$, ${\cal P}$, and $g\in {\bf S}_3$ are utilized.
 We plot examples of $L$ and $V$ dependences of the scaled gap
 $(L/2\pi)\Delta E_{\tilde s}(K,V,L)$ in Fig.\ \ref{FIG1}, and find the
 crossing points. 
 While non-trivial finite-size corrections may affect their behaviors,
 we shall extrapolate them to the thermodynamic limit according to the
 finite-size scaling argument \cite{Derr82}:
 Suppose a single-power formula, i.e., $V_3(K,L)-V_3(K)\propto
 L^{-\psi_3}$, then the exponent is given as $\psi_3=\omega_3+1/\nu$,
 where $\omega_3$ and $\nu$ are the correction exponent and the critical
 exponent of the correlation length.
 For the F three-state Potts model, $\omega_3=\frac45$ \cite{Nien82}
 and $1/\nu=\frac65$, so we shall use the formula with $\psi_3=2$ and
 extrapolate $V_3(K,\bar{L})$ to the limit. 

\begin{figure}[t]
 \includegraphics[width=2.8in]{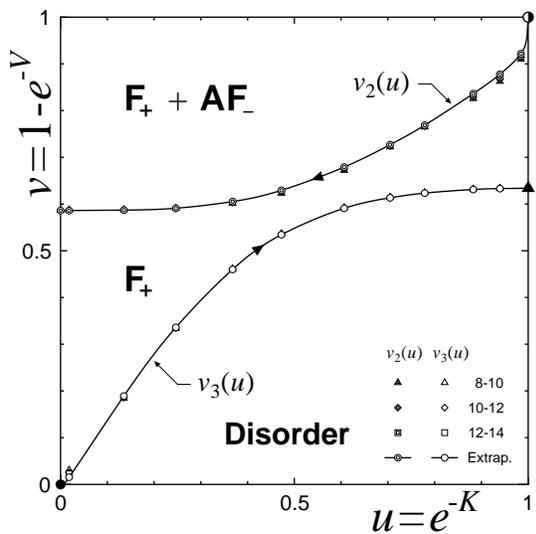}
 \caption{
 The phase diagram of the Hamiltonian\ (\ref{eq_Hami1}) in the space of
 the reduced parameters $(u,v)=(e^{-K},1-e^{-V})$.
 Open circles (double circles) with a curve exhibit the phase boundary
 $v_3(u)$ [$v_2(u)$], which belongs to the three-state Potts (Ising)
 universality class (arrows show the directions of the RG-flows). 
 The filled triangle on the $v$ biaxis denotes the F three-state Potts
 critical point $(1,v^{\rm ex}_3)$, and the filled (half-filled) circle
 corresponds to the ground state of the AF three-state Potts model on
 $\Lambda$ ($\Lambda_-$) with the Gaussian criticality. 
 }
 \label{FIG2}
\end{figure}

 For convenience, we define the reduced couplings
 $(u,v)=(e^{-K},1-e^{-V})$ and compactify the parameter space within a
 unit square region.
 Then we draw the phase boundary line $v_3(u)$ corresponding to
 $V_3(K)$ (open circles with a fitting curve) in Fig.\ \ref{FIG2}. 
 The size-dependent data $V_3(K,\bar{L})$ are also given by other
 marks.
 From this figure, we can find the following: 
 The extrapolated boundary line starts out of the filled circle
 $(u,v)=(0,0)$ {\it linearly} with the increase of $u$, which agrees
 with the crossover argument, i.e., $V_3(K)\propto e^{-K\phi_3}$ with
 $\phi_3=(2-x_P)/(2-x_\epsilon)=1$
 \cite{Delf01,CardText}.
 Then, $v_3(u)$ monotonically increases and finally terminates at the
 filled triangle $(u,v)=(1,v^{\rm ex}_3)$ with $v^{\rm ex}_3:=1-e^{-V^{\rm
 ex}_3}$, as expected.
 Consequently, the boundary line connects the Gaussian and the
 three-state Potts critical fixed points, which is in accord with
 Zamolodchikov's $c$-theorem
 \cite{Zamo86}.

 The line $V_3(K)$ is now known to separate the disordered and the
 partially ordered phases, because the spin degrees on $\Lambda_-$ are
 still disordered, at least, on the decoupling line, $K=0$ and $V>V^{\rm
 ex}_3$.
 Therefore, another phase boundary to the ordered phase exists.
 To see this, let us investigate our model for large enough $V$.
 Since the F order is well established on $\Lambda_+$, we can
 approximately replace the spin degrees on $\Lambda_+$ by a
 spontaneously favored value, e.g., $\tau$.
 Then, the subsystem on $\Lambda_-$ is effectively described by the AF
 three-state Potts model under the uniform magnetic field; the effective
 Hamiltonian is
 \begin{equation}
  {\cal H}_-(\tilde{K},V)
   =\tilde{K}\sum_{j}         \delta_{\sigma_j,\tau}
   +V\sum_{[j,k]} \delta_{\sigma_j,\sigma_{k}}
   \label{eq_Hami2}
 \end{equation}
 ($j,k\in\Lambda_-$). 
 Again, the AF Potts criticality at $(K,V)=(0,\infty)$ is perturbed by
 two relevant operators, but unlike in the case of Eq.\ (\ref{eq_Hami1})
 the ${\bf S}_3$ symmetry is absent, so that a different type of
 crossover may be expected.
 For large $\tilde{K}$, since $\sigma_j$ cannot take $\tau$, the
 subsystem is effectively described by the Ising model.
 R\'acz and Vicsek investigated the Hamiltonian (\ref{eq_Hami2}) by the
 use of the MC method, and obtained the boundary line \cite{Racz83}, so
 that our next numerical task is to determine the second line $V_2(K)$
 to complete the phase diagram of our Hamiltonian\ (\ref{eq_Hami1}).

 The relevant level in our PRG calculations may correspond to the
 magnetic-type excitation on the Ising critical fixed point $\Delta
 E_\sigma(K,V,L)$ with the scaling dimension $x_\sigma=\frac18$,
 and it can be found in the above of the levels which are to be
 degenerate to the ground state.
 So, similarly to $V_3(K,\bar{L})$, we can calculate the size-dependent
 estimates of transition points $V_2(K,\bar{L})$ from the crossings of
 the scaled gaps $L\Delta E_\sigma(K,V,L)$.
 Then, extrapolating them to the thermodynamic limit by using the
 formula
 $V_2(K,L)-V_2(K)\propto L^{-\psi_2}$, with $\psi_2=3$ \cite{Derr82},
 we obtain $V_2(K)$ [the corresponding value $v_2(u)$ is exhibited by
 the double circles with a curve in Fig.\ \ref{FIG2}]. 
 Although Eq.\ (\ref{eq_Hami2}) becomes exact in the limit $V\to\infty$,
 the obtained boundary line is qualitatively comparable to that in Ref.\
 [\onlinecite{Racz83}] also for finite couplings.
 Here, let us examine the limit $V_2(K\to\infty)$. 
 Suppose that the complete F order is established on $\Lambda_+$, then
 the limit is given by the transition point of the Ising model on
 $\Lambda_-$, $V^{\rm ex}_2\simeq0.8814$ ($1-e^{-V^{\rm ex}_2}\simeq
 0.5858$ in Fig.\ \ref{FIG2}).
 However, the spin configurations which are neglected in this argument
 exist in the original Hamiltonian\ (\ref{eq_Hami1}), so $V_2(\infty)$
 may deviate from $V_2^{\rm ex}$.
 Our numerical estimate, $V_2(\infty)\simeq0.8820$, is quite close to
 $V_2^{\rm ex}$.
 This implies that the contributions from the neglected spin
 configurations may be small around $V_2^{\rm ex}$, and thus Eq.\
 (\ref{eq_Hami2}) is expected to provide a very good description even
 near $V_2^{\rm ex}$. 

 Now, in order to evaluate the conformal anomaly numbers, we perform
 three point fitting on Eq.\ (\ref{eq_c}) based on $L-2$, $L$, and
 $L+2$, and estimate the size-dependent values $c_q(L)$ ($q=$2 and 3).
 As an example, we show them at intermediate couplings
 $(K,V)=(1,V_q(1))$ in Table\ \ref{TAB_II}.
 Although nonmonotonic size dependences are visible, the extrapolated
 data agree well with the theoretical values, $\frac12$ and $\frac45$. 
 For parameters close to the IR fixed points, i.e.,
 $(K,V_2(K))$ with $K\gg 1$ and
 $(K,V_3(K))$ with $K\ll 1$,
 there are only small $L$ dependences, and it is easy to confirm the
 criticalities.
 However at $(K,V_q(K))$ far away from the IR fixed points their precise
 estimations become difficult for the present system sizes.

\begin{table}
 \caption{
 The size-dependent estimates of the conformal anomaly numbers $c_q(L)$
 at $(K,V)=(1,V_q(1))$ ($q=$2, 3).
 The fitting
 $c_q(L)=c_q(\infty)+b_1/L^2$
 is performed using the data of $L=$10 and 12.
 }
 \begin{tabular}{cccccc}
  \hline\hline

  ~~~~~~&~~ $L=6$~~ &~~~~~8~~~~~ &~~~~~10~~~~~ &~~~~~12 ~~~~~&  $\infty$ \\ 
  
  \tableline
  $c_2(L)$ &0.5134201~&~0.4359578~&~0.4399469~&~0.4611012~&~0.5092\\
  $c_3(L)$ &0.8012989~&~0.7682470~&~0.7714379~&~0.7815416~&~0.8045\\
  \hline\hline
 \end{tabular}
 \label{TAB_II}
\end{table}


 Finally, we make a remark on the behavior around $(K,V)=(0,\infty)$
 [i.e., the half-filled circle $(u,v)=(1,1)$ in Fig.\ \ref{FIG2}].
 We have considered the model\ (\ref{eq_Hami1}) possessing the global
 ${\bf S}_3$ symmetry.
 With the increase of $V$, ${\bf S}_3$ is broken at $V_3(K)$ and reduced to the
 ${\bf Z}_2$ symmetry with respect to the interchange of unfavored two spin
 states (``charge conjugation'') for $V>V_3(K)$.
 While the lattice Hamiltonian\ (\ref{eq_Hami2}) well describes the
 transition accompanied by the ${\bf Z}_2$ symmetry breaking
 \cite{Racz83,Quei99},
 corresponding field theory in the scaling limit is required to clarify,
 for instance, the phase boundary line around $(K,V)=(0,\infty)$.
 For this issue, it is plausible that the symmetry-breaking negative
 field ($\propto K$) couples with the uniform magnetization $s_j$ and
 brings about a competition to the energy operator $\epsilon_j$ 
 (a coupling $\propto e^{-V}$).
 By borrowing the sine-Gordon expression in Ref.\ [\onlinecite{Delf01}],
 this competition may be formulated by the action preserving the
 sublattice symmetry and the charge conjugation but breaking the cyclic
 permutation symmetry as
 \begin{equation}
  {\cal A}={\cal A}_{c=1}-\int {\rm d}^2x
   \left[
    \mu\cos\beta\varphi+
    \bar\mu\cos\left(\frac{4\pi}{\beta}\tilde{\varphi}\right) 
   \right],
   \label{eq_DSG}
 \end{equation}
 where ${\cal A}_{c=1}$ is the free boson action, and $\tilde{\varphi}$ is the
 dual field of $\varphi$.
 The possibility that the dual sine-Gordon theory\ (\ref{eq_DSG})
 describes the Ising criticality was mentioned by Delfino in relation to
 the Ashkin-Teller model
 \cite{DelfNATO,Delf99}. 
 Further, note that the action\ (\ref{eq_DSG}) can become self-dual at
 $\beta^2=4\pi$ which is known to exactly show the Ising criticality
 \cite{Lech01}.
 Based on this expression, we can predict a shape of the boundary line
 as follows: 
 $K\propto e^{-V_2(K)\phi_2}$ with
 $\phi_2=(2-x_s)/(2-x_\epsilon)=\frac83$.
 This seems to coincide with the {\it rapid} change of the boundary observed
 in Fig.\ \ref{FIG2}.
 However, the precise estimation of $\phi_2$ is outside of the scope of
 our present research; the investigation to confirm this prediction is
 now in progress.


 To summarize, we have investigated two crossovers of criticalities
 in the square-lattice three-state Potts antiferromagnet with staggered
 polarization field. 
 On the basis of the dual sine-Gordon field theories, we have also given
 an argument on the criticalities and the shapes of the phase
 boundaries.
 For more detailed study, we are now performing the Monte Carlo
 simulation calculations to show the evidences for supporting the
 present result; we will give them in the forthcoming article.


 Main computations were performed using the facilities of
 Yukawa Institute for Theoretical Physics. 

\vspace{-4mm}


\begin{references}
 \newcommand{\REF }[4]{#1 {\bf #2}, #3 (#4)}
 \newcommand{\PRL }[3]{\REF{Phys. Rev. Lett.\   }{#1}{#2}{#3}}
 \newcommand{\PRA }[3]{\REF{Phys. Rev.\        A}{#1}{#2}{#3}}
 \newcommand{\PRB }[3]{\REF{Phys. Rev.\        B}{#1}{#2}{#3}}
 \newcommand{\PRD }[3]{\REF{Phys. Rev.\        D}{#1}{#2}{#3}}
 \newcommand{\PRE }[3]{\REF{Phys. Rev.\        E}{#1}{#2}{#3}}
 \newcommand{\PLA }[3]{\REF{Phys. Lett.\       A}{#1}{#2}{#3}}
 \newcommand{\PLB }[3]{\REF{Phys. Lett.\       B}{#1}{#2}{#3}}
 \newcommand{\NPB }[3]{\REF{Nucl. Phys.\       B}{#1}{#2}{#3}}
 \newcommand{\ZPB }[3]{\REF{Z.    Phys.\       B}{#1}{#2}{#3}}
 \newcommand{\CMP }[3]{\REF{Comm. Math. Phys.\  }{#1}{#2}{#3}}
 \newcommand{\JMP }[3]{\REF{J. Math. Phys.\     }{#1}{#2}{#3}}
 \newcommand{\JPSJ}[3]{\REF{J. Phys. Soc. Jpn.\ }{#1}{#2}{#3}}
 \newcommand{\JSP }[3]{\REF{J. Stat. Phys.\     }{#1}{#2}{#3}}
 \newcommand{\JPA }[3]{\REF{J. Phys.\          A}{#1}{#2}{#3}}
 \newcommand{\JPC }[3]{\REF{J. Phys.\          C}{#1}{#2}{#3}}
 \newcommand{\IBID}[3]{\REF{{\it ibid.}}{#1}{#2}{#3}}
 \newcommand{\etal}{{\it et al.}}
\vspace{-4mm}
 \bibitem{Wann50}
 G.H. Wannier, 
 \REF{Phys. Rev.}{79}{357}{1950}. 

 \bibitem{Hout50}
 R.M.F. Houtapple,
 \REF{Physica~(Amsterdam)}{16}{425}{1950}; 
 J. Stephenson,
 \REF{J. Math. Phys. (N.Y.)}{5}{1009}{1964};
 \IBID{11}{420}{1970}. 

 \bibitem{Alex75}
 For example,
 X. Qian, M. Wegewijs, and H.W.J. Bl\"ote,
 \PRE{69}{036127}{2004}, and references therein. 

 \bibitem{Zamo86} 
 A.B. Zamolodchikov,
 \REF{Zh. Eksp. Teor. Fiz.}{43}{565}{1986}
 [\REF{JETP Lett.}{43}{730}{1986}].

 \bibitem{Lena67}
 A. Lenard,
 cited in
 E.H. Lieb,
 \REF{Phys. Rev.}{162}{162}{1967};
 E.H. Lieb, 
 \PRL{18}{692}{1967}. 

 \bibitem{Baxt70}
 R.J. Baxter,
 \REF{J. Math. Phys. (N.Y.)}{11}{3116}{1970};
 R.J. Baxter,
 \REF{Proc. Roy. Soc. London}{A383}{43}{1982}.

 \bibitem{Nijs82}
 M.P.M. den Nijs, M.P. Nightingale, and M. Schick, 
 \PRB{26}{2490}{1982}.

 \bibitem{Burt97}
 J.K. Burton Jr. and C.L. Henley,
 \JPA{30}{8385}{1997}.

 \bibitem{Delf01}
 G. Delfino, 
 \JPA{34}{L311}{2001}. 

 \bibitem{Kola84}
 J. Kolafa,
 \JPA{17}{L777}{1984};
 J.-S. Wang, R.H. Swendsen and R. Koteck\'y, 
 \PRL{63}{109}{1989}.

 \bibitem{Sala98}
 J. Salas and A.D. Sokal,
 \JSP{92}{729}{1998}.

 \bibitem{Card01}
 J.L. Cardy, J.L. Jacobsen, and A. Sokal,
 \JSP{105}{25}{2001}.
 \bibitem{Ferr99}
 S.J. Ferreira and A.D. Sokal,
 \JSP{96}{461}{1999}.

 \bibitem{DelfNATO}
 G. Delfino, 
 hep-th/0110181.

 \bibitem{Lech03}
 P. Lecheminant and E. Orignac,
 \PRB{69}{174409}{2004}.

 \bibitem{Quei02}
 S.L.A. de Queiroz, 
 \PRE{65}{056104}{2002}.

 \bibitem{Lech01}
 P. Lecheminant, A.O. Gogolin, and A.A. Nersesyan, 
 \NPB{639}{502}{2002}.

 \bibitem{Pott52}
 R.B. Potts, 
 \REF{Proc. Cambridge. Philos. Soc.}{48}{106}{1952}.

 \bibitem{Card84}
 J.L. Cardy,
 \JPA{17}{L385}{1984}.

 \bibitem{Blot86}
 H.W. Bl\"ote, J.L. Cardy and M.P. Nightingale,
 \PRL{56}{742}{1986};
 I. Affleck,
 \IBID{56}{746}{1986}.

 \bibitem{OtsuAP}
 For the 1D quantum systems; see
 H. Otsuka,
 \PRB{66}{172411}{2002};
 H. Otsuka and M. Nakamura,
 \PRB{70}{073105}{2004}.
 
 \bibitem{Room80}
 M.P. Nightingale,
 \REF{Physica}{83A}{561}{1976};
 H.H. Roomany and H.W. Wyld, 
 \PRD{21}{3341}{1980}.

 \bibitem{Derr82}
 B. Derrida and L. De. Seze, 
 \REF{J. Phys. (Paris)}{43}{475}{1982}.

 \bibitem{Nien82}
 B. Nienhuis, 
 \JPA{15}{199}{1982}.

 \bibitem{CardText}
 J.L. Cardy, 
 {\em Scaling and Renormalization in Statistical Physics}
 (Cambridge U.P., Cambridge, 1996).

 \bibitem{Racz83}
 Z. R\'acz and T. Vicsek,
 \PRB{27}{2992}{1983}. 

 \bibitem{Quei99}
 S.L.A. de Queiroz \etal,
 \PRE{59}{2772}{1999}.

 \bibitem{Delf99}
 G. Delfino, 
 \PLB{450}{196}{1999}. 
  
\end{references}
\end{document}